\documentclass{aa}
\usepackage{graphicx}
\usepackage{hyperref}

\begin{document}

\def\gsim{\!\!\!\phantom{\ge}\smash{\buildrel{}\over
  {\lower2.5dd\hbox{$\buildrel{\lower2dd\hbox{$\displaystyle>$}}\over
                               \sim$}}}\,\,}

\def\lsim{\!\!\!\phantom{\le}\smash{\buildrel{}\over
  {\lower2.5dd\hbox{$\buildrel{\lower2dd\hbox{$\displaystyle<$}}\over
                               \sim$}}}\,\,}
\def\kms{\rm ~km~s$^{-1}$}

  \title{High Resolution Spectroscopy of Balmer-Dominated Shocks in the RCW 86, Kepler and SN~1006 Supernova Remnants
\thanks{Based on observations collected at the European Southern Observatory, 
Paranal, Chile (ESO Programme 67.D-0579).}}

   \author{
Jesper Sollerman,\inst{1}
Parviz Ghavamian,\inst{2,4}
Peter Lundqvist,\inst{1}
R. Chris Smith\inst{3,4}
}

   \offprints{jesper@astro.su.se}

\institute{Stockholm Observatory, Department of Astronomy, AlbaNova, 
SE-106 91 Stockholm, Sweden  
\and 
Department of Physics and Astronomy, Rutgers University, 
136 Frelinghuysen
Rd., Piscataway, NJ 08854-8019
\and
Cerro Tololo Inter-American Observatory, Casilla 603, Chile 
\and
Visiting Astronomer, Cerro Tololo Inter-American Observatory, National Optical Astronomy
Observatories.  CTIO is operated by AURA, Inc. under contract to the National Science Foundation.
}

   \date{}

\authorrunning {Sollerman et al.}
\titlerunning{The collisionless shocks in supernova remnants}

\abstract{ We report results from high resolution optical spectroscopy
of three non-radiative galactic supernova remnants, RCW 86, Kepler's
supernova remnant and SN 1006.  We have measured the narrow component
H$\alpha$ line widths in Balmer-dominated filaments in RCW 86 and  SN
1006, as well as the narrow component width in a Balmer-dominated knot
in Kepler's SNR.  The narrow component line widths measured in RCW 86
and Kepler's SNR show FWHM of  $30-40$\kms, similar to what has been
seen in other Balmer-dominated remnants.  Of the remnants in our
sample, SN 1006 is the fastest shock ($\sim$3000\kms).  
The narrow component H$\alpha$ and H$\beta$ lines in this remnant have a
FWHM of merely 21\kms. Comparing the narrow component widths
measured in our sample with those measured in other remnants shows
that the  width of the narrow component does not correlate in a simple
way with the shock velocity.  The implications for the pre-heating
mechanism responsible for the observed line widths are discussed.

}

\maketitle              

\section{Introduction}

Supernova (SN) explosions produce some of the strongest shocks in
nature.  At the low ISM densities (n\,$\leq\,$1 cm$^{-3}$) such shocks
are often non-radiative, losing  a negligible fraction of their
internal energy to radiative cooling.  
The low
density of the preshock gas makes Coulomb collisions between charged
particle species very infrequent.  
The slow subsequent
Coulomb heating in fast shocks ensures unequal temperatures of
electrons and ions in the postshock flow.

Unlike a radiative shock, the optical emission from a non-radiative 
shock is produced entirely by collisional excitation in the ionization zone 
immediately behind the shock (Raymond 1991).  If the preshock medium is 
significantly neutral, the emission from
a non-radiative shock is dominated by hydrogen line emission 
(i.e., it is Balmer-dominated, Chevalier \& Raymond 1978).  
The Balmer line profiles are quite remarkable.  When a non-radiative shock 
encounters partially neutral gas, cold
hydrogen atoms overrun by the shock are collisionally excited. 
The radiative decay of these excited neutrals produces narrow-component Balmer
emission with a line width determined by the pre-shock temperature.  
On the other hand, charge exchange between cold neutrals and protons
produces fast hydrogen atoms with the velocity distribution of the
postshock protons.  Collisional excitation of these fast neutrals
produces broad Balmer line emission 
(Chevalier \& Raymond 1978; Chevalier et al. 1980).

Balmer-dominated shocks have been detected in five Galactic supernova
remnants (SNRs): SN 1006 and Tycho's SNR 
(Kirshner et al. 1987; Smith et al. 1991; Ghavamian et al. 2002),  
Kepler's SNR  (Blair et al. 1991), 
RCW 86 (Long \& Blair 1990; Smith 1997; Ghavamian et al. 2001) 
and the Cygnus Loop
(Raymond et al. 1983; Hester et al. 1994; Ghavamian et al. 2001).
Four remnants in the Large Magellanic Cloud also exhibit
Balmer-dominated emission (Tuohy et al. 1982; Smith et al. 1991, 1994).  
The width of the broad component Balmer line and the
broad-to-narrow flux ratio observed in the spectra of these remnants
can be used to simultaneously measure the shock speed and the ratio
T$_{\rm e}$/T$_{\rm i}$ immediately behind the shock 
(e.g., Ghavamian et al. 2001).

During the last ten years, high-resolution spectra of Balmer dominated
shocks have revealed narrow component line widths ranging from 
30\kms\ to 50\kms, implying preshock temperatures of 20,000$-$60,000~K
(Hester et al. 1994; Smith et al. 1994).  This is clearly
larger than the thermal width of ambient, undisturbed interstellar
matter.  If the narrow component widths  represent thermal broadening,
their large values in Balmer-dominated SNRs suggests
that the ambient gas
is somehow heated in a precursor by energetic particles which escape
upstream and deposit their energy into the preshock gas 
(Hester et al. 1994; Smith et al. 1994).  These particles may be
cosmic rays, and such precursors are predicted by  diffusive shock
acceleration models~(\cite{be87}).  Alternatively, the same fast
neutrals responsible for the broad component H$\alpha$ emission 
could escape upstream and deposit their energy
into the preshock gas via exchange and elastic collisions, thus forming a fast
neutral precursor (Smith et al. 1994; Lim \& Raga 1996).

\begin{figure*}[!t]
\setlength{\unitlength}{1mm}
\begin{picture}(185,65)(0,0)
\put (0, 0){\includegraphics[width=56mm,bb=167 244 445 549,clip]{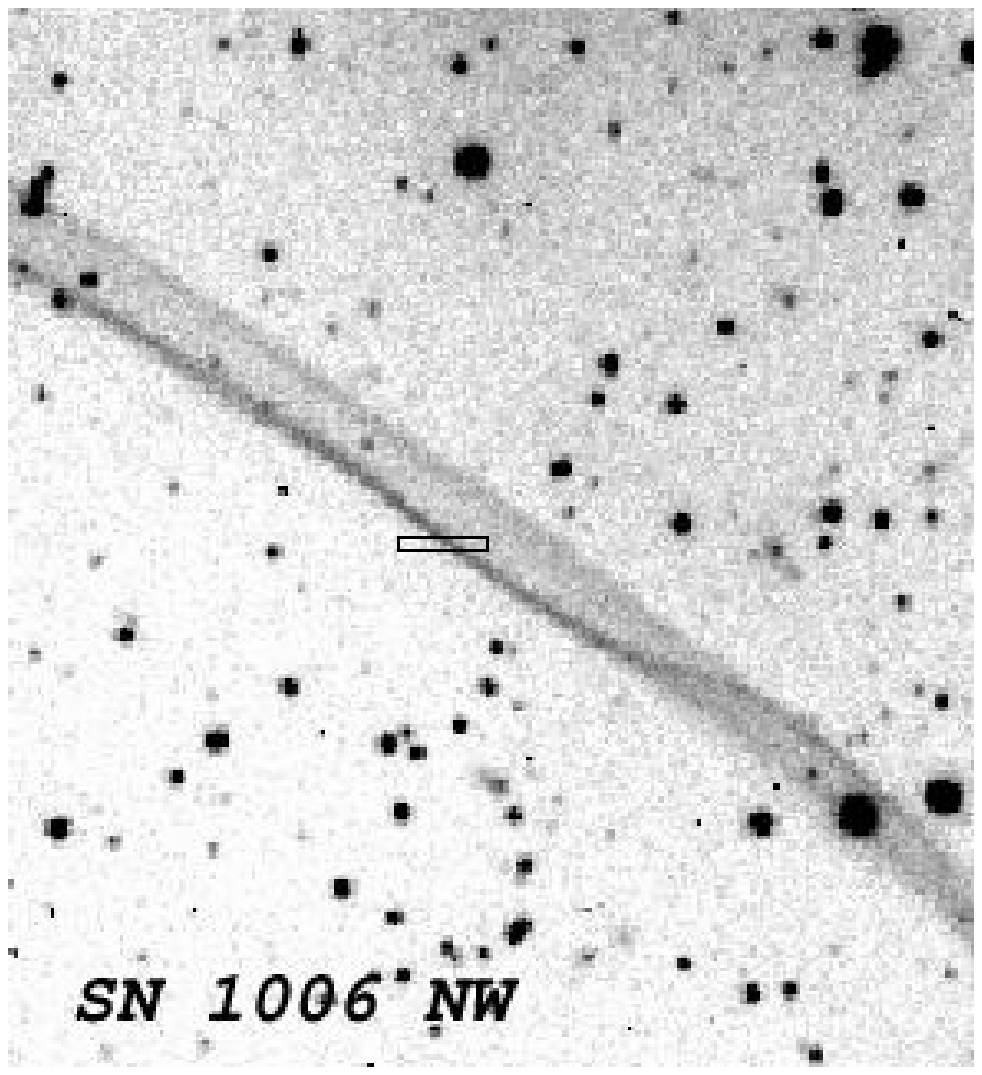}}
\put (58, 0){\includegraphics[width=60mm,bb=43 43 361.551 377,clip]{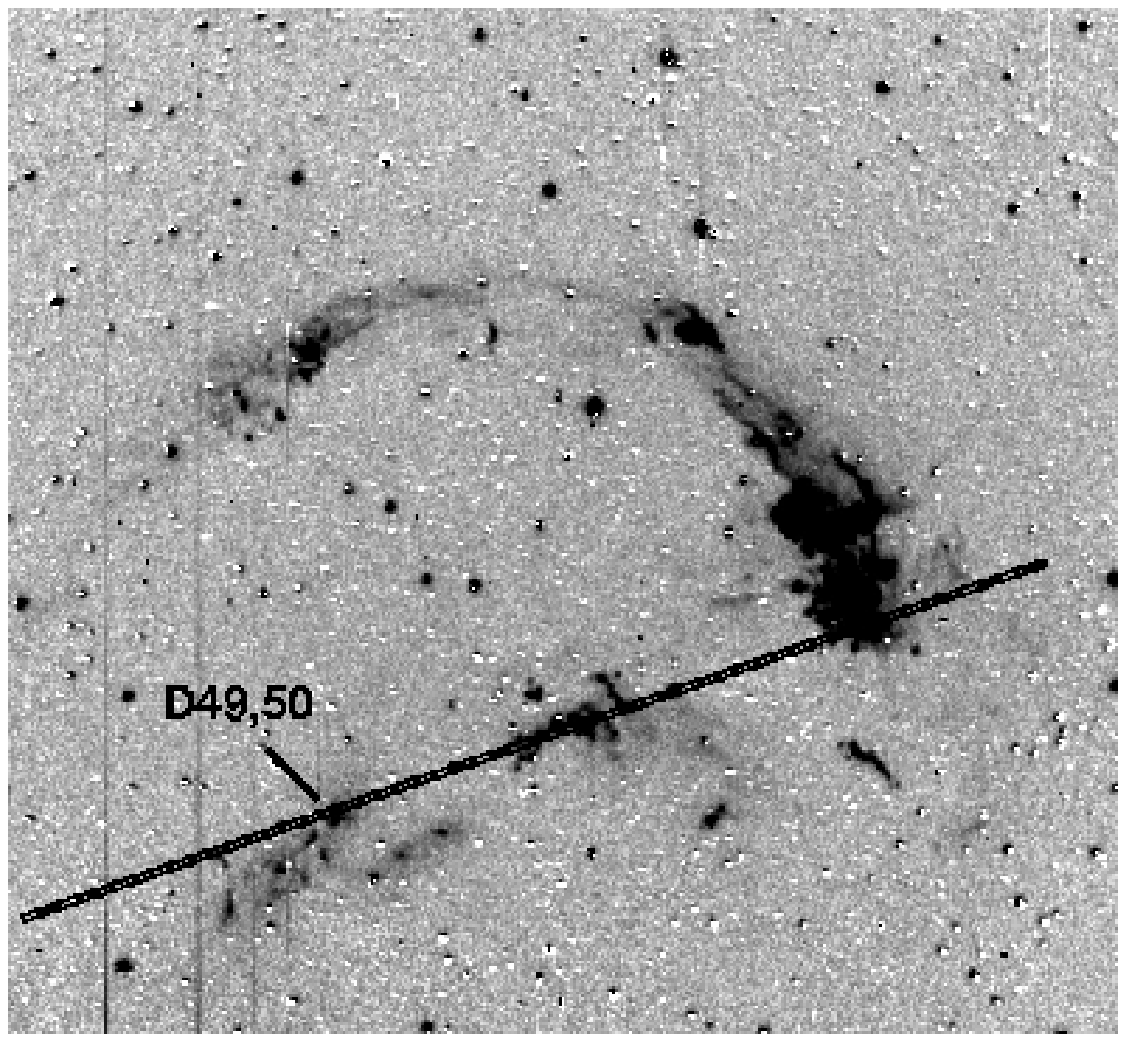}}
\put (120, 0){\includegraphics[width=60mm,bb=0 0 368 344,clip]{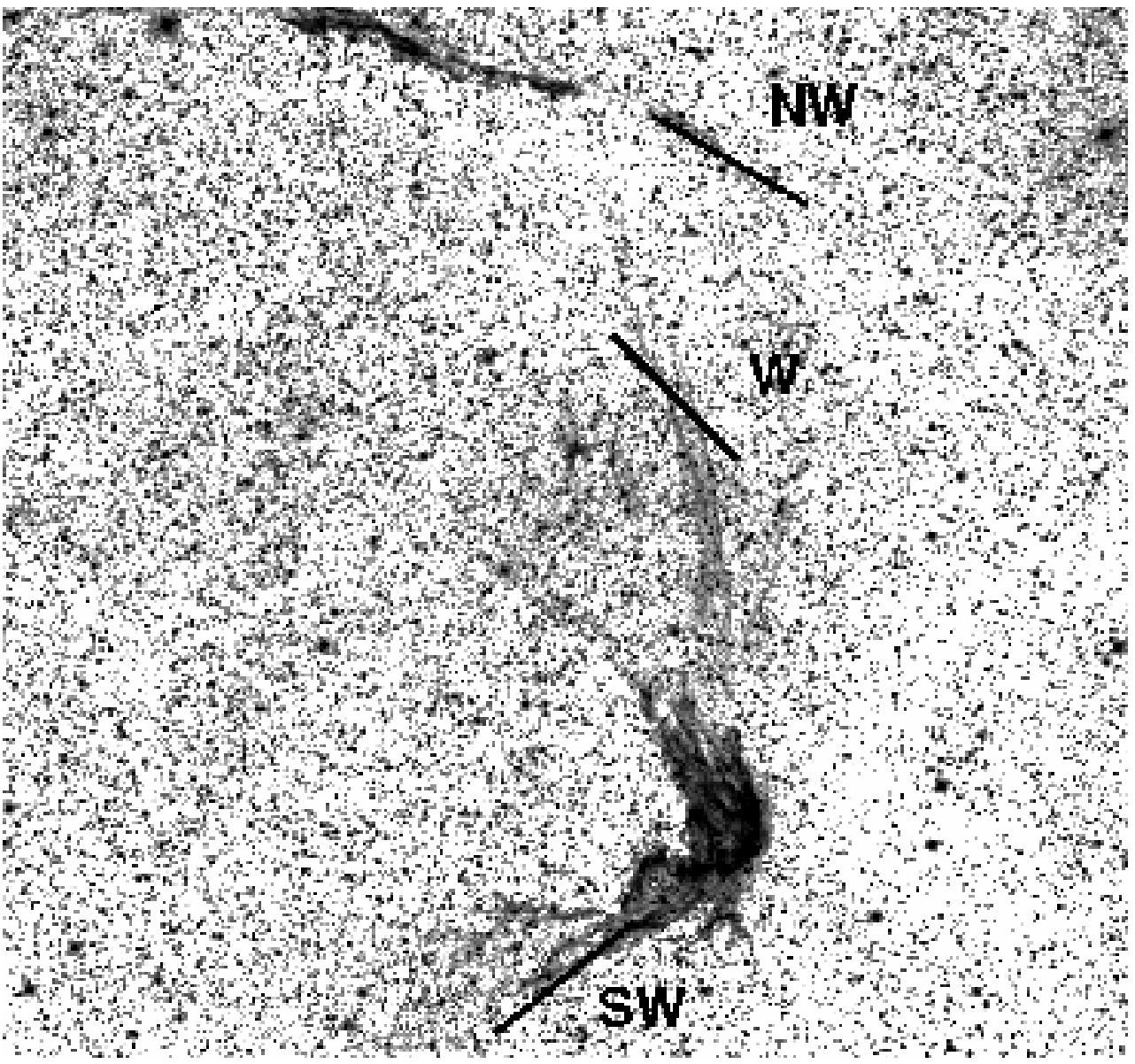}}
\end{picture}
\caption{The positions of the slits used for our observations of the 
three supernova remnants. The leftmost image 
(from Winkler \& Long~1997) shows SN 1006 with the 
12\arcsec\ UVES slit indicated.
North is up and east to the left in all three images.
The middle image (from Blair et al. 1991) shows the Kepler supernova remnant, 
indicating our slit position and the knots D49 \& D50. The slit length is 
5\farcm68. The rightmost image (from Smith 1997) 
shows our three slit-positions used for RCW 86. Slit lengths are 
again 5\farcm68.}
\end{figure*}

The correlation between the shock velocity and the width of the narrow
component line could be a potentially important discriminant between
different precursor scenarios.  
A preliminary study
of upstream heating by fast neutrals suggests that the degree of
precursor heating should be sensitive to the shock speed, degree of
electron-ion temperature equilibration at the shock front and the
preshock neutral fraction (Smith et al. 1994).
On the other hand, it is not clear whether 
heating of the upstream gas by a cosmic ray precursor
should have a strong dependence on the shock velocity
(e.g., Hester et al. 1994; Smith et al. 1994; Boulares \& Cox 1988).

In this investigation we have obtained the first high resolution
optical spectra of three Balmer-dominated Galactic supernova
remnants, RCW 86, Kepler's supernova remnant and SN 1006.
In Sect.~2 we present our observations and the techniques
used to reduce the data.  In Sect.~3 we present the obtained results
and in Sect.~4 we discuss these  results in conjunction with other
data on nonradiative supernova shocks.  Finally, the results are
summarized in Sect.~5.

\section{Observations and data reduction}

\subsection{RCW 86 and Kepler}

The galactic SNR RCW 86 was observed on 1999 June 7, using the
Echelle spectrograph at the f/7.8 focus of the 4~m telescope of Cerro
Tololo Inter-American Observatory.  The T2K detector was connected to
the Red Long Camera and binned by two in the spatial dimension, giving
a plate scale of 0\farcs53 pixel$^{-1}$.  The decker was set to the open
position, giving a spectrograph slit length of 5\farcm68.  However,
vignetting limited the useful length of slit to around 3\farcm68.  
The 79 lines mm$^{-1}$ Echelle grating was used for the observations, with the
cross disperser replaced by a flat mirror.  The width of the
spectrograph slit was 1\farcs3, which combined with the grating
arrangement yielded a spectral resolution of 11.2\kms.  The
spectra were centered on the H$\alpha$ line, while all other emission lines
were excluded by inserting an H$\alpha$ filter 
($\lambda_{\rm c}$=6563 \AA , $\Delta \lambda$=75 \AA ). 
Wavelength calibration was performed
using ThAr lamp spectra interleaved with object exposure frames.

The spectroscopic observations of RCW 86 targeted 
three different Balmer-dominated
filaments of the remnant (Fig.~1).
The atmospheric seeing during the RCW 86 observations ranged from 
1\farcs5 to 2\arcsec.
Of these filaments, the brightest is the one located in the
southwestern region of the remnant, where the Balmer-dominated shocks
lie close to bright radiative shocks (Rosado et al. 1996; Smith 1997;
Ghavamian et al. 2001).  Filaments located in the NW and W portions of RCW 86
are nearly twice as faint as those in the SW.  
However, moderate resolution spectra
show that the widths of the broad component H$\alpha$ lines are 
comparable in all
three regions, ranging from 560\kms, in the SW (Ghavamian et al. 2001) to 
around 600\kms, in the NW/W (Ghavamian 1999).  
The procedures for targeting the Balmer-dominated shocks
and acquisition of the Echelle spectra were as follows:

(1) RCW 86 NW - The spectrograph slit was centered on a star
located at $\alpha$=14:40:23.6, $\delta=-$62:16:46.9 (2000), then offset 
1$\arcsec$ S
and rotated to PA = 240$^{\circ}$.  Five frames were acquired at this
position for a total integration time of 4500 s.

(2)  RCW 86 W - The spectrograph slit was centered on a star
located at $\alpha$=14:40:38.8, $\delta=-$62:24:35.4 (2000), then offset 
1$\arcsec$ S
and rotated to PA = 225$^{\circ}$.  Five frames were acquired at this
position for a total integration time of 4500 s.

(3) RCW 86 SW - The spectrograph slit was centered on a star
located at $\alpha$=14:41:09.4, $\delta=-$62:43:57.0 (2000), then rotated 
to PA = 308.7$^{\circ}$.  Five frames were acquired at this position for a
total integration time of 3000 s.

High resolution spectra of Kepler's SNR were obtained on the same night as
the RCW 86 observations.  During this time the seeing had degraded, ranging
between 3\arcsec\ and 4\arcsec.  The Kepler observations targeted the knotty 
structures in the remnant interior identified as Knots 
D49 \& D50 (D'Odorico et al. 1986; Blair et al. 1991).  
The slit was oriented at PA = 290$^{\circ}$ and placed so 
that the upper portion of the slit
intersected the bright radiative knots along the NW part of Kepler's
SNR, with fainter radiative knots seen near the middle of the slit and
the pure Balmer-dominated Knots D49 \& D50 seen in the lower half
of the slit (Fig.~1).  The spectrograph slit was centered on the
position $\alpha$=17:30:40.5, $\delta=-$21:29:23.0 (2000), 
then rotated to PA = 290$^{\circ}$.  
Five frames were acquired at this position for a total
integration time of 4500 s.

The RCW 86 and Kepler's SNR data were reduced in IRAF using the standard 
processes of
overscan and bias correction, dark count correction and flat fielding.
Wavelength solutions were computed at evenly spaced intervals along
the slit and used to untilt the dispersion axes of the spectra.
Finally, the background was subtracted from each two-dimensional
spectrum using sky emission adjacent to each target along the slit.

\begin{figure}[t]
\begin{center}
\includegraphics[width=.45\textwidth]{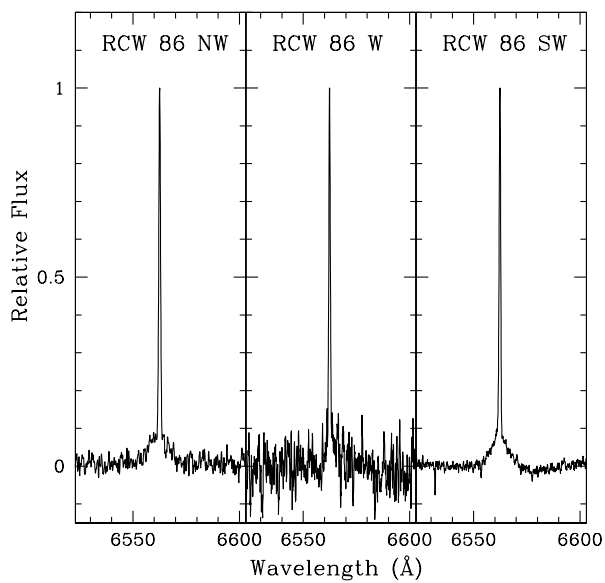}
\end{center}
\caption[]{One-dimensional sky-subtracted spectra of the NW, W and SW 
Balmer-dominated shocks in RCW 86.  The middle spectrum has been smoothed
over 0.8 \AA, while the NW and SW spectra have been smoothed over 0.4 \AA.  
The broad component is clearly detected in the NW and SW spectra, 
with the faint W spectrum exhibiting weak evidence of a broad component.}
\label{rcw1d}
\end{figure}

\subsection{SN 1006}

The observations of SN 1006 were obtained with the
Ultraviolet and Visual Echelle Spectrograph (UVES).
This instrument is located
at one Nasmyth focus on the second unit telescope (Kueyen) of
the Very Large Telescope (VLT) on Paranal, Chile. 
UVES is a high-resolution two-arm cross-dispersed Echelle spectrograph, 
where both arms can be operated separately or simultaneously using a 
dichroic beam-splitter.

The VLT/UVES observations were conducted in visitor mode on June 13, 2001.  
As we were primarily interested in H$\alpha$ and H$\beta$ we chose to 
observe in  the red arm only. With Cross-disperser $\#$3 and the order 
sorter filter SHP700 this set-up\footnote{See
{http://www.eso.org/instruments/uves/} for details on the 
instrument and observational technique.}, 
centered on 5800~\AA, covers the wavelength region 
$4760-6840$~\AA.

UVES has a short slit with a length of 12$^{\prime\prime}$. 
We used a slit width of $1\farcs8$ positioned as
shown in Fig.~1.  The red arm is equipped with a mosaic of two
2k$\times$4k CCDs, with a scale of $0.022-0.026$~\AA~pixel$^{-1}$ and a
spatial scale of $0\farcs18$
pixel$^{-1}$.
The position shown in Fig.~1 was observed in 7 exposures of 15 minutes
each, in total 6300 s. 
The airmass was secz=$1.16-1.48$ and the
seeing varied between $1\farcs1$ and $1\farcs5$.

The spectra were interactively reduced using the 
UVES-pipeline\footnote{{http://www.eso.org/observing/dfo/quality/}}
(version 1.2.0) 
as implemented in MIDAS.  
The reduction separates the
two CCDs and performs bias subtraction and flat-fielding of the data
using calibration frames obtained in the morning.

Wavelength calibration was done by comparison to ThAr arc lamps.  More
than 1000 spectral lines were fitted for both of the CCDs, with an RMS
of better than 0.008~\AA. The resolution  for arc-lines with the
$1\farcs8$ 
slit is $\sim$13\kms, or 0.29~\AA~at the position
of H$\alpha$. 

The pipeline finally extracts the spectrum of the reduced frames.  We
averaged the signal over
5\farcs5 along the slit, 
and subtracted the skylines from the region at
the very end of the slit.

\section{Results}

\begin{figure*}[!t]
\begin{center}
\includegraphics[width=.8\textwidth, angle=0]{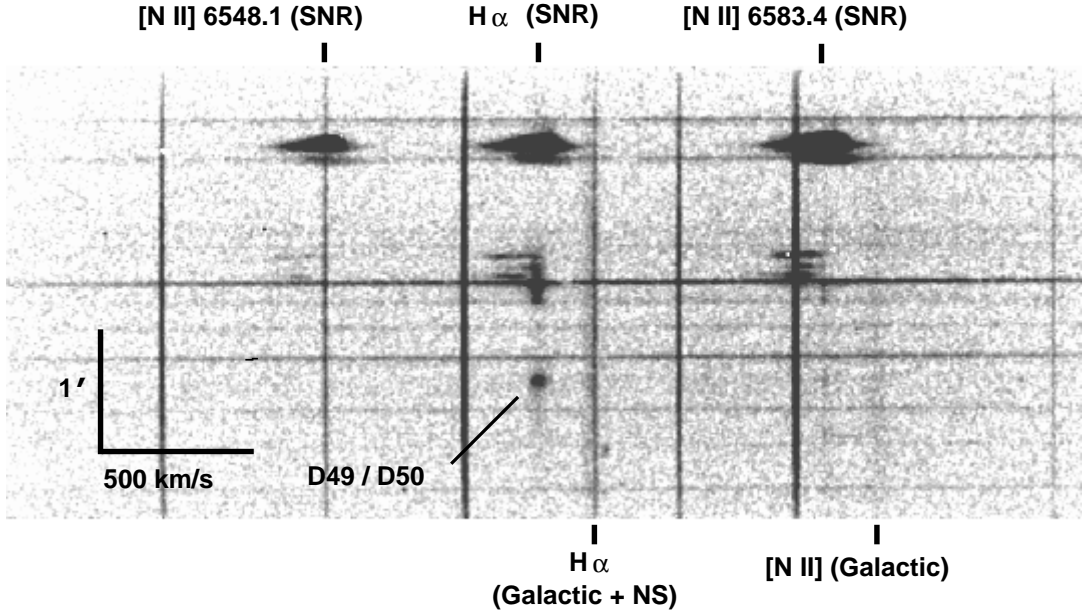}
\end{center}
\caption[]{
The two-dimensional spectrum of Kepler's SNR, shown smoothed with a 1 pixel 
Gaussian and without subtraction of night sky lines.  
The Balmer-dominated knots D49 \& D50 are marked,
along with the corresponding [N~II] emission lines from the SNR and the 
Galactic background.
The remaining bright lines are night sky (NS) emission features.  
The teardrop shaped structures at
the top of the slit and the fainter structures near the middle of the 
slit are radiative shocks
in clumpy, nitrogen-rich material associated with Kepler's SNR.}
\label{kepler2d}
\end{figure*}

\subsection{RCW 86}

The spectra of all three observed positions in RCW 86 appear in
Fig.~2.  The NW and W spectra  were obtained by summing 15$\arcsec$
and 10$\arcsec$ of emission from each filament respectively,  while
the emission from the bright SW filament was summed over 2\farcs5.
The broad component is 
detected in all three spectra.
Although a two-component fit was performed for each of the spectra,
the narrow component line in each case is bright and well separated in
width from the broad component.  This makes the estimated narrow
component width rather insensitive to the inclusion of a second component in
the fit. The measured narrow component FWHM widths were
40$\pm$2\kms, 32$\pm$5\kms\ and 32$\pm$2\kms\ for the 
NW, W and SW regions, respectively.  Due to the large
dispersion of the Echelle spectra, the broad component wings are not
well detected, and our fits yield only lower limits on the broad component 
widths. The fitted values are $\gsim$450\kms, consistent with 
measurements at lower resolution (Long \& Blair 1990; Ghavamian 1999; 
Ghavamian et al. 2001).

\subsection{Kepler's SNR}

In the two-dimensional spectrum of Kepler's SNR (Fig.~3) a number of
knots appear along the length of the slit exhibiting H$\alpha$ and
[N~II] line emission. 
The bottom knot 
corresponds to knots D49 \& D50 tabulated by D'Odorico et al. (1986)
and identified as
Balmer-dominated shocks by Blair et al. (1991).   
Most of the
structures seen along the Echelle slit exhibit a systemic blueshift of
185\kms.

In Fig.~4 we present the
spectrum of D49 \& D50 with the night sky lines included to
illustrate the large width of the H$\alpha$ narrow component.  The
width of the narrow H$\alpha$ component in knots D49 \& D50 is 
42$\pm$3\kms\ after correction for instrumental broadening.  The faint,
broad hump of emission observed in Fig.~4 is broad component
H$\alpha$ emission (v$_{\rm FWHM}\,\approx$1800\kms\ according to
Blair et al. 1991) from knots D49 \& D50. 
At the high resolution of these observations, it is difficult to 
measure the parameters of this component precisely, but both the 
centroid and broad-to-narrow flux ratio are consistent with those 
reported in Blair et al. (1991; $\sim6571$ \AA~and 1.2, respectively)

The non-sky subtracted spectrum of knots D49 \& D50 (Fig.~4) exhibits
two faint [N~II] $\lambda$6583 features, one seen at close to zero
velocity and one blueshifted to 185\kms, the systemic speed of
Kepler's SNR.  The line at zero velocity is rather broad ($\sim$100\kms); 
the two-dimensional spectrum (Fig.~3) shows that this line is
of nearly uniform surface brightness and is seen along the full length
of the unvignetted spectrograph slit.  The most likely origin for the
broad [N~II] is Galactic emission.  Its large width is likely due to a
blend of multiple [N~II] lines from clouds along the line of sight.

More interesting is the narrow, blueshifted [N~II]
$\lambda$6583 line seen in the spectrum of knots D49 \& D50.  From the
two-dimensional spectrum (Fig.~3) it is clear that this [N~II]
emission arises 
from knots D49 \& D50.  
After correction for instrumental
broadening, we find a width of $20\pm$2\kms\ for the [N~II] $\lambda$6583
line in knots D49 \& D50.   
The presence of this line in 2,000\kms\ Balmer-dominated shocks
is further discussed below (Sect. 4.3).

\begin{figure}[]
\begin{center}
\includegraphics[width=.5\textwidth]{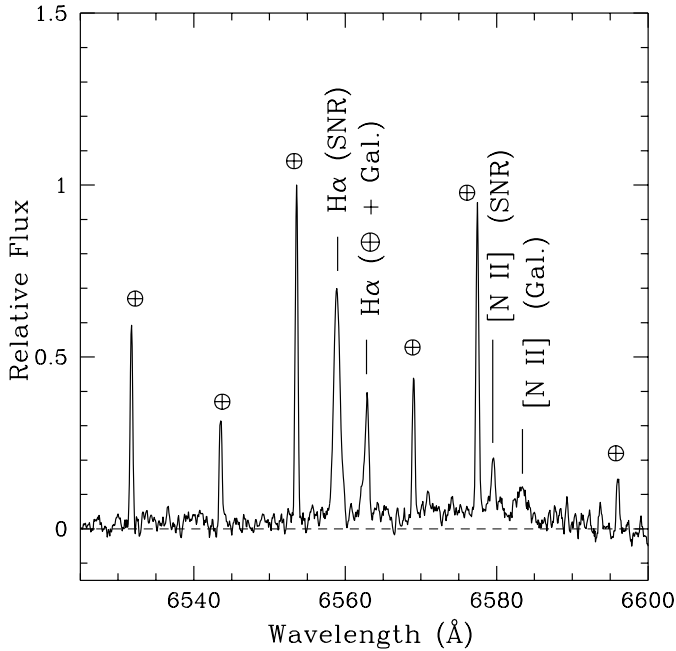}
\end{center}
\caption{The extracted spectrum of Balmer-dominated knots D49 \& D50 in 
Kepler's SNR, smoothed over 0.4 \AA.  Night sky lines have been left in 
and are marked.  There are three emission lines associated
with the knots: narrow H$\alpha$ (42\kms) and broad H$\alpha$ 
($\sim$2000\kms) lines and
a narrow [N~II] $\lambda$6583 line (20\kms).  
All three lines are blueshifted 185\kms\ from zero velocity.
The unshifted, broad [N~II] line seen near 6583 \AA\, 
is Galactic emission (see discussion in text).}
\label{kepler1d}
\end{figure}

\subsection{SN 1006}

In Fig.~5 we show the fully reduced H$\alpha$ line from the part of 
SN~1006 shown in Fig.~1. 
The narrow Balmer line
is indeed resolved. 
We measure the FWHM by a Gaussian fit to be 24\kms. This is
clearly broader than the surrounding skylines, which show a FWHM of
about 12\kms. Thus, correcting the width of the H$\alpha$ line
for the instrumental resolution, we arrive at an intrinsic width of
21\kms. The inset in Fig.~5 shows a Gaussian fit 10$\%$ wider 
and narrower, and from this we estimate the error in the narrow 
component FWHM to be $\lsim$3\kms.

\begin{figure}
\begin{center}
\includegraphics[width=.55\textwidth]{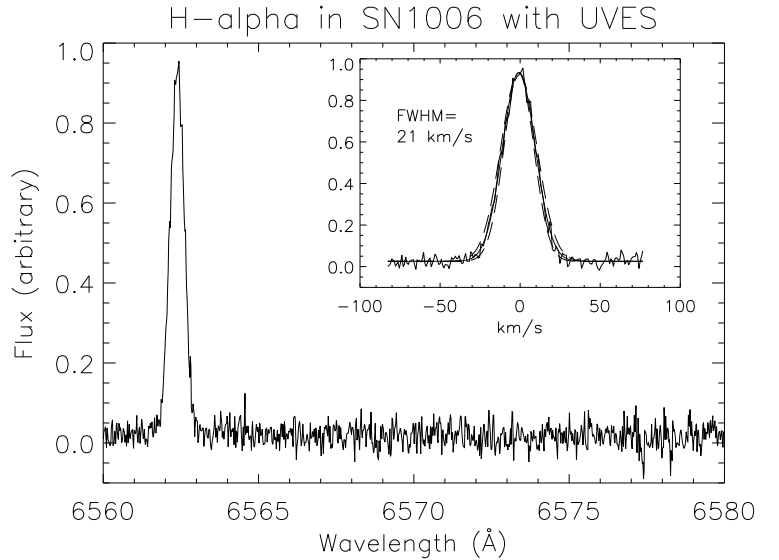}
\end{center}
\caption[]{The narrow H$\alpha$ line detected in SN~1006. The inset shows 
the line in velocity-space, with the best Gaussian fit of 
FWHM=21\kms\ (full line). The broken lines indicate Gaussian fits 
with FWHM which are 10$\%$ wider and narrower. Zero velocity refers to the
6562.4 \AA. The broad component of H$\alpha$ (cf. Ghavamian et al.
2002) is not detected in these high resolution observations.}
\label{eps2}
\end{figure}

Apart from H$\alpha$, and consistent with the medium-resolution spectra of
Ghavamian et al. (2002), the only other line detected in SN 1006 is H$\beta$. 
While this line is more noisy, the width can nevertheless be measured by
a Gaussian fit.  The width corrected for instrumental resolution 
is the same as for H$\alpha$, i.e., 21\kms.
The nearby skylines have widths of slightly more than 12\kms, 
consistent with the resolution obtained in the arc lamp spectral lines.

\section{Analysis and Discussion}

\subsection{Comparison to other Balmer dominated supernova remnants}

The narrow component in several other Balmer dominated shocks have been 
spectroscopically resolved. 
Smith et al.~(1994) studied four remnants in the LMC, 
and found narrow component widths of $30-50$\kms. 
The same year, Hester et al.~(1994) investigated the Balmer dominated
parts of the Cygnus loop and found similar values for the narrow
components. 
We have included our five narrow component width measurements 
in Table~1, along with a listing of prior measurements in other SNRs for
comparison.

The shocks in Table~1 of RCW 86, Cygnus and Tycho were modeled in
detail by  Ghavamian et al. (2001), 
and SN 1006 by Ghavamian et al. (2002). 
The shock velocities for the three LMC remnants in Table~1
were estimated by Raymond~(2001)
based on the observations by Smith et al.~(1991)
and extrapolating the relation between shock velocity and
equilibration.

\begin{table*}[ht]
\begin{center}
\caption[]{Balmer dominated supernova remnants}
\begin{tabular}{lcccc}
\hline
\noalign{\smallskip}

SNR & Shock velocity & Narrow component & References &  $\kappa$ \\
    &  (km s$^{-1}$) & FWHM (km s$^{-1}$) & & (cm$^{2}$ s$^{-1}$)    \\

\noalign{\smallskip}
\hline
\noalign{\smallskip}

Cygnus Loop  & $300-400$ & $28-35$ & GRSH01,HRB94 &\\
RCW 86 SW & $580-660$     & $32\pm2$      & GRSH01,This work & $<$ 3.3$\times$10$^{26}$/n$_{\rm e}$\\
RCW 86 W  & $580-660$      & $32\pm5$      & This work &  $<$ 1.2$\times$10$^{27}$/n$_{\rm e}$ \\
RCW 86 NW & $580-660$      & $40\pm2$      & This work &  $<$ 1.5$\times$10$^{25}$/n$_{\rm e}$ \\
Kepler D49 \& D50 & $2000-2500$ & $42\pm3$ & BLV91,This work&  $<$ 1.7$\times$10$^{27}$/n$_{\rm e}$ \\
0505-67.9       & $440-880$       & $32-43$ & SKBW91,R01,SRL94 &\\
0548-70.4       & $700-950$       & $32-58$ & SKBW91,R01,SRL94 & \\
0519-69.0       & $1100-1500$     & $39-42$ & SKBW91,R01,SRL94 &\\
0509-67.5       & $--$          &   $25-31$ & SRL94 &\\
Tycho   & $1940-2300$     & $44\pm4$      & GRSH01,GRHB00 &\\
SN 1006 & $2890\pm100$  & $21\pm3$      & GWRL02, This work &\\

\noalign{\smallskip}
\hline
\end{tabular}
\label{TABLE:}
\end{center}
\end{table*}

The broad component width measured for knots D49 \& D50 in Kepler's supernova 
remnant by Blair et al. (1991) 
is 1700$-$2000\kms, similar to the value measured for Knot g in Tycho's SNR 
(Smith et al. 1991; Ghavamian et al. 2001).  
However, the broad-to-narrow ratio for D49 \& D50 is 1.2, nearly twice as
large as that of Tycho Knot g.  
Based on our non-radiative shock models (Ghavamian et al. 2001, 2002) we
estimate that a shock speed of 2000$-$2500\kms\ is appropriate for Kepler 
knots D49 \& D50. This is assuming a low electron-ion equilibration at 
the shock front, T$_{\rm e}$/T$_{\rm p}$ $<$ 0.2.
For the Cygnus Loop it was not exactly the same region of the SNR measured 
for the broad and narrow component.
For RCW 86, the broad component widths measured from low resolution 
spectra are nearly 
identical in all three regions (Ghavamian 1999; Ghavamian et al. 2001), 
so we have assumed that the shock speeds are also equal.

We see that the widths of the narrow components 
determined for RCW 86 and Kepler are very similar to those
found in other supernova remnants. In fact, the main conclusion to be drawn 
from Table~1 is that the
narrow component widths remain remarkably constant over a factor of 10 in 
estimated shock velocity. The widths
we measure for RCW 86, Kepler's SNR and SN 1006 strengthen this conclusion.   
The noticeable drop in 
narrow component width at the highest shock speed, that of SN 1006, 
is still only a factor of $\sim2$ lower than the largest widths measured 
thus far.

\begin{figure}
\begin{center}
\includegraphics[width=.55\textwidth]{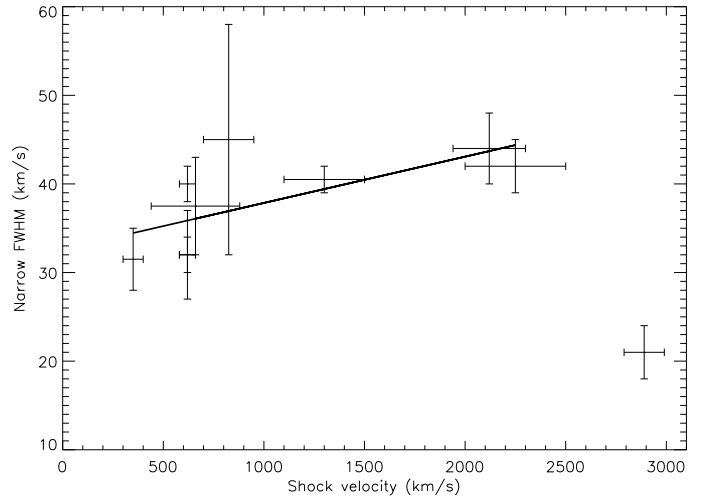}
\end{center}
\caption[]{Shock velocity versus the FWHM of the narrow component of the 
observations listed in Table~1, except for 0509-67.5 for which there is no
estimated shock velocity. The line shows the best linear fit to the data, 
as discussed in the text.}
\label{correlation}
\end{figure}

In Fig.~6 we have plotted the data from Table~1. 
If we disregard the data from SN~1006 
(discussed below), a weak trend for broader narrow components in faster 
shocks can be seen. A simple linear chi-square  ($\chi^{2}$) 
fit that only 
allows for 
errors in the narrow component FWHM gives a slope of 
5.2($\pm1.6$)\kms\ for every 1000\kms\ increase in shock 
velocity. This fit has a reduced $\chi^{2}$ of 1.6. A constant line does 
not give a good fit. It gives a reduced $\chi^{2}$ of 2.7, which is 
rejected at a $2.8\sigma$ level. 
Thus if these error bars are to be interpreted formally (and if SN 1006
can be ignored), a preshock heating mechanism with modest but positive
velocity dependence is favored.
It remains to be seen if detailed modeling of cosmic ray pre-heating 
(e.g., Boulares \& Cox 1988) can account for this trend.

However, the alternative scenario may be even more problematic.
Though some of the broad component neutrals undoubtedly cross 
into the preshock region, their energy distribution and number density
are sensitive to a number of parameters. 
A neutral precursor model that favors a strong 
(roughly quadratic, Smith et al. 1994)
dependence on the shock velocity is clearly not consistent
with the relatively small range of narrow component
widths among the Balmer-dominated remnants summarized in Table~1. 
This was noted already by Smith et al. (1994).

\subsection{The narrow component line width in SN 1006}

As seen in Table~1 and in Fig.~6, the narrowest of the lines is the one 
in SN 1006.
The resolved narrow component of the
Balmer lines has a width of merely 21\kms. 
This corresponds to a kinetic temperature of $\sim9500$~K. 
The shock velocity of SN 1006 has recently been accurately determined when a 
deep spectrum of SN 1006 was obtained by Ghavamian et al. (2002). 
Summing up emission along 51$^{\prime\prime}$ of their slit, 
they were able to detect even He~I $\lambda$6678. 
These observations, together with a detailed non-radiative shock model, 
confirmed a very low degree of electron-proton equilibration at the shock 
front, $T_{\rm e}/T_{\rm p}\leq$ 0.07, and 
the shock speed was determined to be 
2890$\pm$100\kms.

Our data thus indicate that the fastest of the Balmer-dominated
shocks in our sample displays a line
width close to that expected for the undisturbed, warm component of the ISM 
(Ferri\'ere 2001; Raymond 2001).
Whatever mechanism broadens the narrow component in slower shocks is
likely to operate also at higher Mach numbers. But these data do not
favor a model where the pre-heating is simply correlated with the shock
velocity, as suggested for fast neutrals 
(Smith et al. 1994, but see also Hester et al. 1994). 

The process heating the pre-shock gas in SN~1006 appears to
be less efficient than in the other SNRs listed in Table~1. 
Detailed modeling of the energy deposition of cosmic rays or neutrals passing 
upstream through the shock is still in its infancy 
(e.g., Lim \& Raga 1996; Boulares \& Coc 1988) 
and is outside the scope of this paper.  

It is unclear whether our result for the SN 1006 Balmer filaments
reflects a general trend of sharply declining narrow component line
width at high shock speeds ($\simeq$\,3000\kms) or whether the
small line width is caused by some unique, unknown property of the SN
1006 blast wave.

It is interesting to note that also SNR 0509$-$67.5 displays a rather
narrow component (Table~1).  In fact, Smith et al. (1994) measured
25$\pm$3\kms\ for the central and eastern rims.  This is the one LMC
remnant whose broad component H$\alpha$ width has eluded detection
(Tuohy et al. 1982; Smith et al. 1991; Smith et al. 1994).
Interestingly, the limit on the blast wave speed derived by Tuohy et
al. (1982) for 0509$-$67.5 is $\geq\,$3600\kms, i.e., similar to
the speed of 3000\kms,  measured for SN~1006. 
In fact, speculations of the
narrow component width in this remnant were already forwarded by Smith
et al. (1994).  However, it is probably premature to use these small
widths of 0509$-$67.5  in conjunction with our SN~1006 results to
argue for a high shock velocity  cutoff in narrow component line
widths in Balmer-dominated shocks. In the western rim of the same
supernova remnant,  Smith et al. (1994) measure a narrow component
width of 31$\pm$2\kms, which is similar to the values seen in many
other Balmer-dominated remnants.   A more reliable estimate of the
blast wave speed of 0509$-$67.5 will be  required to confirm the
presence of a high shock velocity cutoff.  Measurements of the width
of the narrow component of  H$\alpha$ at more positions within SN~1006
could also be useful, to  investigate whether the narrow width is
sensitive to geometry or local pre-shock conditions.

\subsection{Narrow [N~II] emission in the Balmer dominated Knots in Kepler's SNR}

In the case of Kepler's SNR, the detection of [N~II] from
Balmer-dominated shocks may have some bearing on the precursor
question. With a broad component H$\alpha$ width of 1800\kms 
(Blair et al. 1991), the inferred shock speed in the knots is far too 
high for the onset of postshock cooling.  
Therefore, the [N~II] emission from knots
D49 \& D50 cannot be produced in a cooling zone as is seen in
partially radiative shocks in RCW 86 (Long \& Blair 1990).  Further
evidence against a cooling zone origin for the [N~II] is the lack of
[S~II] and [O~III] emission  from knots D49 \& D50 (Blair et al. 1991).  
The width of the [N~II] line observed from Knots D49 \& D50 indicates that the
nitrogen emission cannot come from the immediate postshock ionization
zone either, since for shock speeds $\sim$2000\kms, nitrogen ions
would be raised to temperatures exceeding 5$\times$10$^{7}$~K by
collisionless heating at the shock front.

The remaining possibility is that the observed [N~II] emission is
generated ahead of the Balmer-dominated shocks in knots D49 \& D50, in
the precursor.  The difference between the narrow component H$\alpha$
width and the [N~II] line width is interesting.  If the [N~II] line
width is thermal, it would indicate that the nitrogen ions ahead of
the shock are heated to $\sim$120,000~K, nearly three times hotter
than the hydrogen atoms.  This would be difficult to explain with a
fast neutral precursor because in that case the energy is deposited
primarily into the H atoms.  In addition, there is little time for
Coulomb collisions to communicate the heating to the N ions before
they are all overtaken by the shock.  Therefore, it is likely that the
20\kms\ broadening of the [N~II] line is nonthermal.  A nonthermal
broadening mechanism is readily found in the Alfv\`{e}n wave activity
expected in a cosmic ray precursor.  Such waves follow a spectrum of
amplitudes (Blandford \& Eichler 1987; Smith et al. 1994) which can
simultaneously heat ions with a range of masses.  Therefore, the large
[N~II] line width in knots D49 \& D50 may be indirect evidence of a
cosmic ray precursor.  Note that this does not disprove the existence
of a fast neutral precursor, but implies that such a mechanism alone
is unlikely to account  for the observations.  The fact that [N~II] is
detected here but not in Balmer-dominated shocks of comparable speeds
in other SNRs (such as Tycho's SNR, Smith et al. 1991; Ghavamian et
al. 2001) may be due to an overabundance of nitrogen inferred for the
knotty material around Kepler's SNR 
(Leibowitz \& Danziger 1983; Blair et al. 1991).

\subsection{Limits on the diffusion coefficient of precursor particles}

The narrow component line widths measured in RCW 86 and Kepler's SNR
can be used to place a limit on the cosmic ray diffusion coefficient $\kappa$ 
in these remnants.  Assuming that the line widths represent thermal
broadening, an upper limit on the precursor width is obtained by
requiring that it must be small enough to avoid complete ionization of
neutral H before it enters the shock (Hester et al. 1994;
Smith et al. 1994). We have refrained from making the same analysis for the 
fast shock in SN 1006, since for a preshock temperature of $\sim$9500 K
the ionization argument can be questioned.
The limit on the diffusion coefficient is then
$\kappa\,<\,$v$_{\rm S}^{2}$/n$_{\rm e}$q$_{\rm i}$(T$_{\rm e}$), 
where n$_{\rm e}$ is the
preshock electron density and q$_{\rm i}$(T$_{\rm e}$) is the ionization rate
coefficient at the preshock electron temperature T$_{\rm e}$.   We have
computed the limits on $\kappa$ using the fits of Janev et al. (1987)
to q$_{\rm i}$, the shock velocity estimates from Blair et al.
(1991) for the Balmer-dominated knots in Kepler's SNR, and the shock
velocity estimates of Ghavamian et al. (2001) for
RCW 86. The results appear in Table~1. Variations in the preshock
electron density due to variations in preshock ionization fraction and
total density can cause variations in $\kappa$ by an order of
magnitude.  However, the diffusion coefficients derived here are
smaller than the value  10$^{29}$ cm$^{2}$ s$^{-1}$ expected for the
galactic cosmic ray population (Ginzburg \& Syrovatskii 1964).
Therefore, the coefficients derived here are at least consistent with
the existence of a cosmic ray population confined to a region close to
the shock front by Alfv\'{e}n wave turbulence.

\subsection{The Kinematic Distance of RCW 86}

One source of recent debate regarding RCW 86 has been the possible
association of the remnant with a neighboring OB stellar association.
Rosado et al. (1996) performed Fabry-Perot spectroscopy of the  bright 
radiative shocks in the SW portion of the remnant.  They used the fitted line
profiles to  estimate the systemic velocity of RCW 86, and then used
Galactic rotation curve parameters to estimate the distance to the
remnant.  
We can redo the same analysis.
In fact, our high spectroscopic resolution
allows us to better constrain the velocity centroid of the narrow
component. Moreover, the neutral gas producing the narrow H$\alpha$ line
is not affected by the passage of the blast wave shock 
(e.g., Chevalier \& Raymond 1978), so the measured line center
more accurately reflects the systemic velocity of the local ambient
gas than the line center measured from the
radiative shocks.  
Following Rosado et al. (1996) we convert the
observed radial velocity to a velocity relative to the LSR, then use
V$_{\rm LSR}$ to compute a kinematic distance to RCW 86 from the fits of
Brand \& Blitz (1993) to the Galactic  rotation curve.  In our
analysis we assume that the galactocentric distance R$_{0}\,=\,$8.5
kpc for the Sun and that the angular speed 
$\omega_{0}\,=\,$25.9\kms\ kpc$^{-1}$ for the solar neighborhood (Brand \& Blitz 1993).

Using the line profile fits for the SW, W and NW blast wave spectra of
RCW 86, we obtained the radial velocity from each spectrum using
velocity centroids of the narrow component lines.  From the
wavelengths of the night sky OH and H$\alpha$ lines in the 1-D
spectra, we estimate that our measured radial velocities for the
narrow blast wave H$\alpha$ lines were accurate to within 2\kms.
Transforming the velocities to LSR values, we find that
V$_{\rm LSR}=-$33$\pm$3\kms, for the SW, W and NW shocks.  Using the
rotation curves of Brand \& Blitz (1993), we find that the distance to
RCW 86 implied by V$_{\rm LSR}$ in the 
Balmer-dominated shocks is 2.3$\pm$0.2 kpc.
This is in excellent agreement with the value of
V$_{\rm LSR}$=$-$33.2$\pm$4.5\kms, determined by Rosado et al. (1996), 
and within
the errors our distance estimate of 2.3$\pm$0.2 kpc agrees with the
value of 2.8$\pm$0.4 from Rosado et al. (1996).  A distance of
2.3$\pm$0.2 kpc places RCW 86 in close proximity to the OB association
stars catalogued by Lyng\aa~(1964).  From photometry of stars in the
association, Westerlund (1969) and Moffat \& Cameron (1999) estimated
a distance of 2$-$2.5 kpc for the association.

\section{Summary}

We have added five more measurements of the narrow components in 
non-radiative supernova shocks. 
The measurements in RCW 86 and in Kepler's SNR show FWHM of 
$\sim 30-40$\kms, similar to what is seen in other non-radiative
Balmer-dominated shocks. 
The extended sample show only a weak trend for increasing FWHM of 
the narrow lines with increasing shock velocity. The exception is the 
fastest shock in our sample, that in the NW limb of SN 1006.  
The resolved H$\alpha$ and H$\beta$ 
lines have a FWHM of only  $21\pm3$\kms. This is
lower than measured in any other Balmer dominated supernova remnant, and
disfavors any precursor mechanism which simply correlates the shock velocity
with the degree of pre-heating. 

We have also made a first measurement of the narrow component width in
a Balmer-dominated knot (Kepler's SNR), where we have also detected
faint [N~II] $\lambda$6583 emission.  The [N~II] line width is
significantly larger than expected for thermal broadening.  We argue
that the [N~II] line width is caused by broadening from Alfv\`{e}n
wave turbulence in a cosmic ray precursor. 

Finally, we have estimated a distance of 2.3$\pm$0.2 kpc to RCW 86, 
which places it at the same position as a
known OB association, supporting a core collapse SN origin for RCW 86.

\begin{acknowledgements}
We thank the referee, K. Long, for useful comments on the manuscript,
and J. Raymond,  D. Cox and S. Reynolds for discussions
This work was supported by the Swedish Research Council. PG acknowledges 
support from Chandra Grant GO0-1035X. PL is a Research Fellow at the Royal 
Swedish Academy supported by a grant from the Wallenberg Foundation.
\end{acknowledgements}

\end{document}